\documentclass[superscriptaddress,twocolumn,aps,prl]{revtex4-1}
\usepackage{natbib}
\usepackage[utf8]{inputenc}
\usepackage{amsmath,amssymb,color}
\usepackage{graphicx}
\usepackage{latexsym}
\usepackage{multirow}

\begin{document}

\title{The scaling of exploding liquid jets under intense X-ray pulses}
\author{ Alfonso M. Ga\~n\'an-Calvo}
\affiliation{Departamento de Ingenier{\'\i}a Aerospacial y Mec\'anica de Fluidos, ETSI, Universidad de Sevilla.\\
Camino de los Descubrimientos s/n 41092, Spain.}

\begin{abstract}
A general scaling of the evolution of an exploding liquid jet under an ultra short and intense X-ray pulse from a X-ray free electron laser (XFEL) is proposed. A general formulation of the conservation of energy for blasts in vacuum partially against a deformable object leads to a compact expression that governs the evolution of the gap produced by the explosion. The theoretical analysis contemplates two asymptotic stages for small and large times from the initiation of the blast. A complete dimensional analysis of the problem and an optimal collapse of experimental data reveal that the universal approximate analytical solution proposed is in remarkable agreement with experiments.
\end{abstract}

\pacs{47.55.D-, 47.55.db, 47.55.df}

\maketitle

The damage exerted by blasts in partial contact with deformable objects has been the subject of diverse studies of interest in mining \cite{Wei2009}, defense \cite{He2010}, or forensic applications \cite{Goldstein2012,Wightman2001}. However, probably the strongest scientific interest in a detailed description of the damage caused by intense blasts has recently come with the advent of serial femtosecond crystallography (SFX) \cite{Cetal11}. In this application, a microscopic jet produced by Flow Focusing \cite{G98a,DWSWSSD08} and loaded with protein crystal samples is exposed to focused femtosecond X-ray pulses from XFELs with energies in the range of mJ. As a consequence, the jet explodes in a vacuum chamber, generating a blast that splits the jet in two portions. These explosions were admirably reported in detail by \cite{Stan2016} from large series of experiments. The use of increasing frequencies of pulsation of XFELs up to the MHz range \cite{Wiedorn2018} to maximize the sample hit ratio has challenged the SFX practitioners. The damage caused to the samples upstream of the blast has been the subject of increasing attention \cite{Grunbein2018}. In this work we focus on the evolution of the blast \cite{Stan2016} to precisely describe the physics of the process, obtaining the maximum expansion velocities and stresses undergone by the liquid column and the energies involved. To do that, we first propose a general formulation of the problem of blasts in vacuum, partially against deformable objects.

\begin{figure}[htb]
\centering
\includegraphics[width=0.45\textwidth]{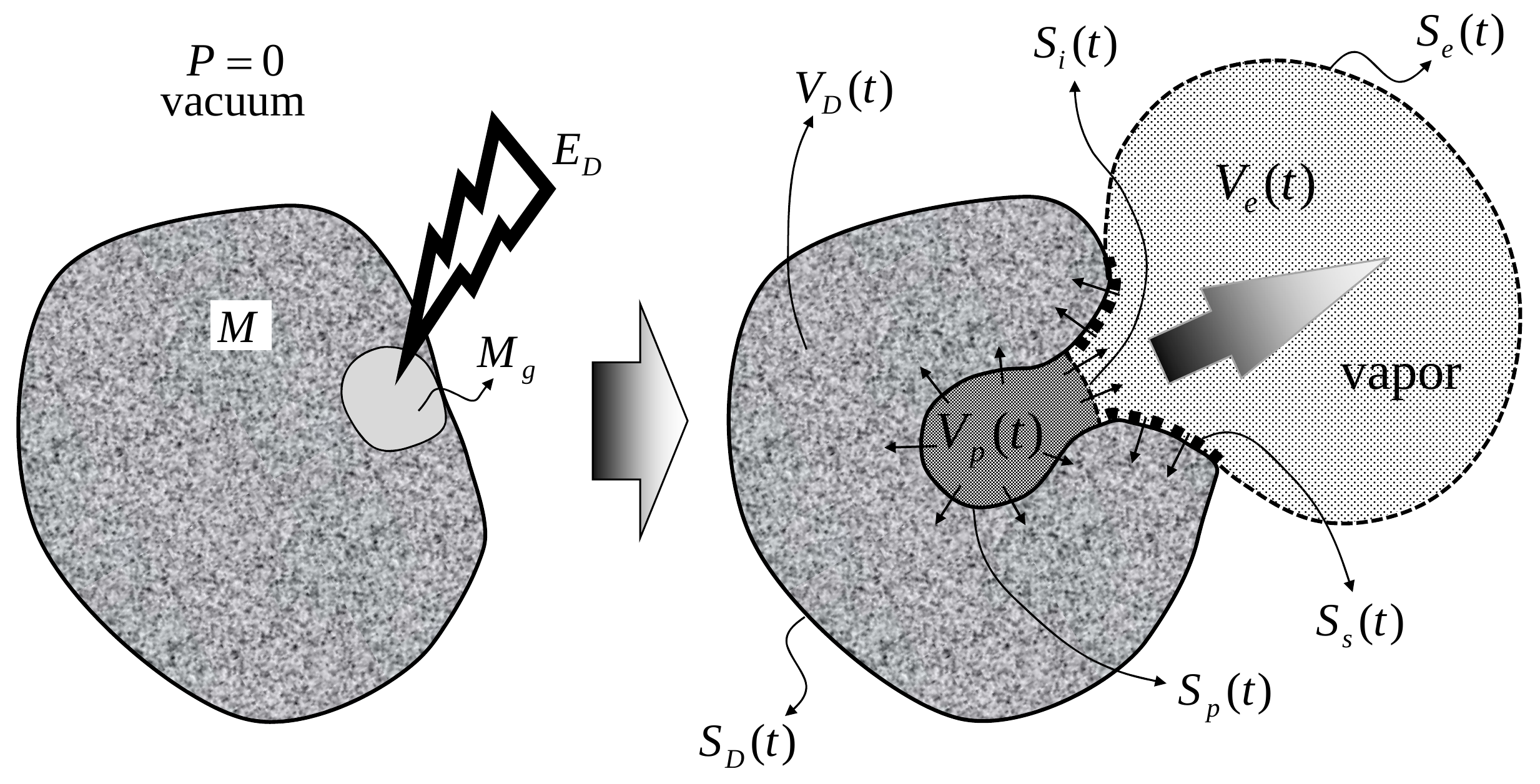}
\vspace{-3mm}
\caption{General sketch of the problem. The small arrows at $S_i$, $S_s$, and $S_p$ indicate the expected direction of the motion of said surfaces. Thick dotted lines indicate $S_s(t)$}
\label{fig1a}\vspace{-2mm}
\end{figure}

\begin{figure}[htb]
\centering
\includegraphics[width=0.340\textwidth]{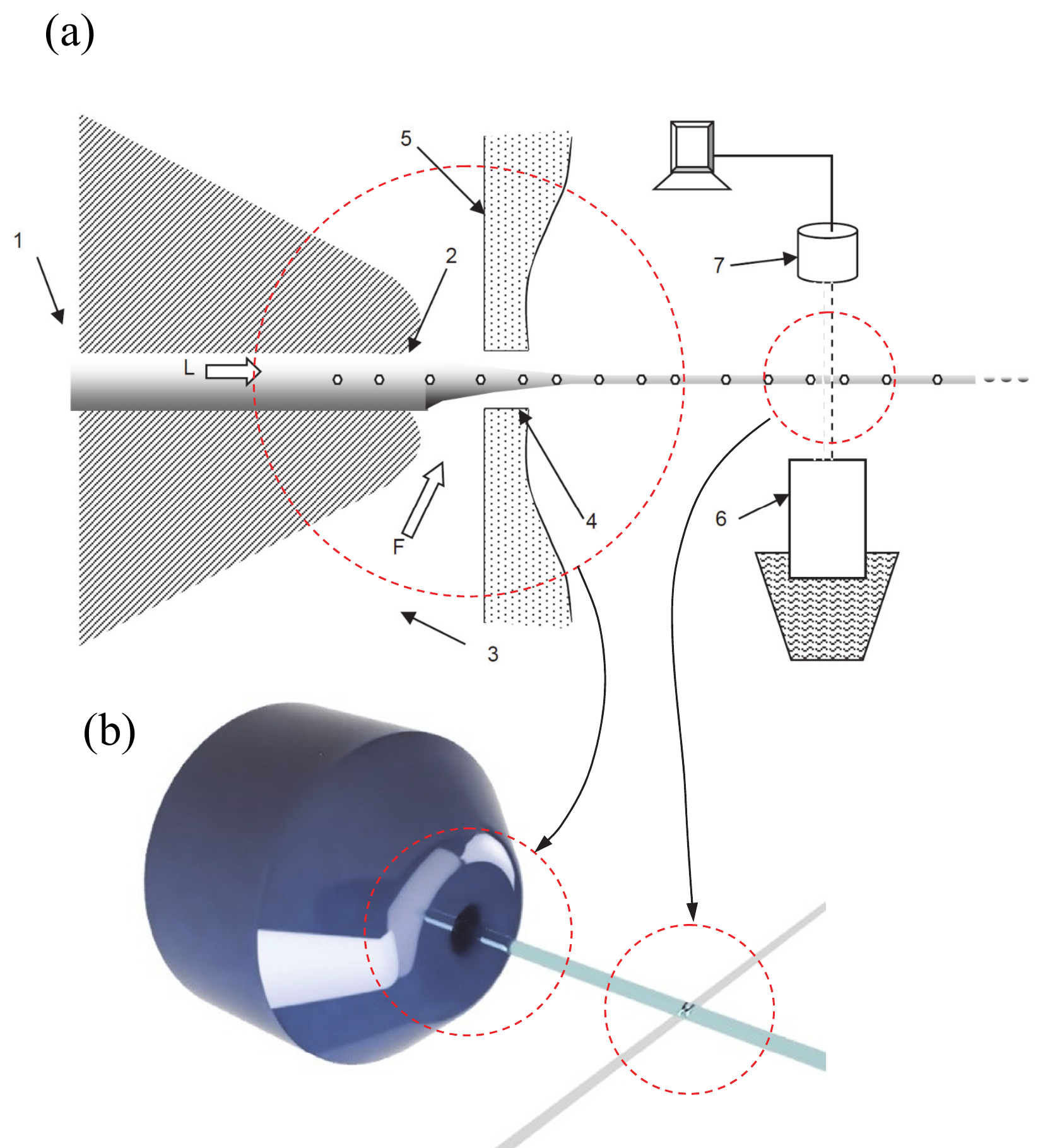}
\vspace{-3mm}
\caption{(a) Original sketch in Ga{\~n\'a}n-Calvo et al 2003 \cite{GananCalvo2003}. 1: Liquid carrying samples (e.g. protein microcrystals); 2: Liquid jet issued from the transport capillary due to focusing gas F (Helium in current experiments in XFELs); 3: Virtual gas focusing nozzle; 4: Discharge orifice; 5: Focusing element (e.g. a plate); 6: Interrogating beam source; 7: Detector. (b) Sketch of the nozzle-jet-beam configuration used in the X-ray experiments (courtesy B. Ga{\~n\'a}n-Riesco, Ingeniatrics Tec.).}\label{fig0}\vspace{-2mm}
\end{figure}

{\it General compact formulation of blasts in partial contact with deformable objects.-} Consider a finite amount of condensed deformable matter $M$ surrounded by vacuum. An amount of energy $E_D$ largely sufficient to vaporize a portion $M_g$ of that matter is suddenly injected in it. If part of $M_g$ is adjacent or in direct contact with the outer surface of $M$, it violently expands as a blast both into vacuum and against $M$ (figure \ref{fig1a}). The expanding hot gas domain $V_g(t)$ can be consider to comprise two virtual subdomains formally defined in the Supplemental Material: (i) $V_p(t)$, that pushes and deforms $M$ by dominant pressure forces, and (ii) $V_e(t)$, whose mass and energy are constant by definition while it expands into vacuum. Therefore, $V_e(t)$ does not make any work on $M$. Those definitions and their implications do not impose any artificial condition on the natural evolution of the total gas domain $V_g(t)$. However, they provide a drastic simplification: defining $V_o=V_p(t_o)+V_e(t_o)$ as the initial energized volume at the initial instant $t_o$, the fraction $\chi=V_p(t_o)/V_o$ is a problem parameter since it is constant along the process by definition.
Given that the analysis of $V_e$ is irrelevant since its energy is constant and its evolution is decoupled from $M$, one can write the following compact equation of conservation of energy that governs the coupled evolution of $M$ and $V_p$:
\begin{equation}
\int_{t_o}^t\int_{S_p+S_i}P\, \mathbf{v}\cdot \mathbf{n}\text{d} A\text{d} t' +\frac{P_oV_p(t)^{1-\gamma}V_o^\gamma}{(\gamma -1)}=\chi E_D
\label{ep1}
\end{equation}
where the first term in the left side is the {\it total} work made by $V_p$ on $M$ since the beginning of the blast, and the second is the internal energy left in $V_p$ at a given $t$. The initial conditions are $V_p(t_o)=\chi V_o$, $P(t_o)=P_o$, and $P_o V_o/(\gamma-1)=E_D$. Since the initial value of the internal energy is the same for $V_p$ and $V_e$, $\chi$ is also the injected energy fraction contained in $V_p$ at the beginning of the blast, or the {\it efficiency} of the blast against $M$. The factor $(\gamma-1)$ can be considered the Gr\"uneisen coefficient of the initial energized matter (in many cases a {\it warm dense matter state} \cite{Beyerlein2018}), $\gamma$ being its adiabatic coefficient when it expands. Consistently with the assumption made in \cite{Stan2016} for water irradiated by an intense ultrashort X-ray pulse, if one makes the usual assumption that the evolution of the gas in the blast is quasi-isentropic, this coefficient stays nearly constant from large to small densities along the blast, as we will see in the analysis of experimental results. The first term in the right hand side of equation (\ref{ep1}) is the total energy received by the object from the start of the blast up to time $t$. This compact formulation is particularly advantageous for relatively simple geometries of deformable objects that allow the explicit expression of that first term in terms of a single geometrical variable. This is the case of the explosion of microjets in SFX \cite{Cetal11,Stan2016}, where a capillary liquid microjet produced by flow focusing \cite{G98a} gently carries the samples in series as initially suggested in \cite{GananCalvo2003} (figure \ref{fig0}). Here, the microjet is shot by a train of extremely short X-ray pulses \cite{Cetal11,Wiedorn2018}.

\begin{figure}[htb]
\centering
\includegraphics[width=0.35\textwidth]{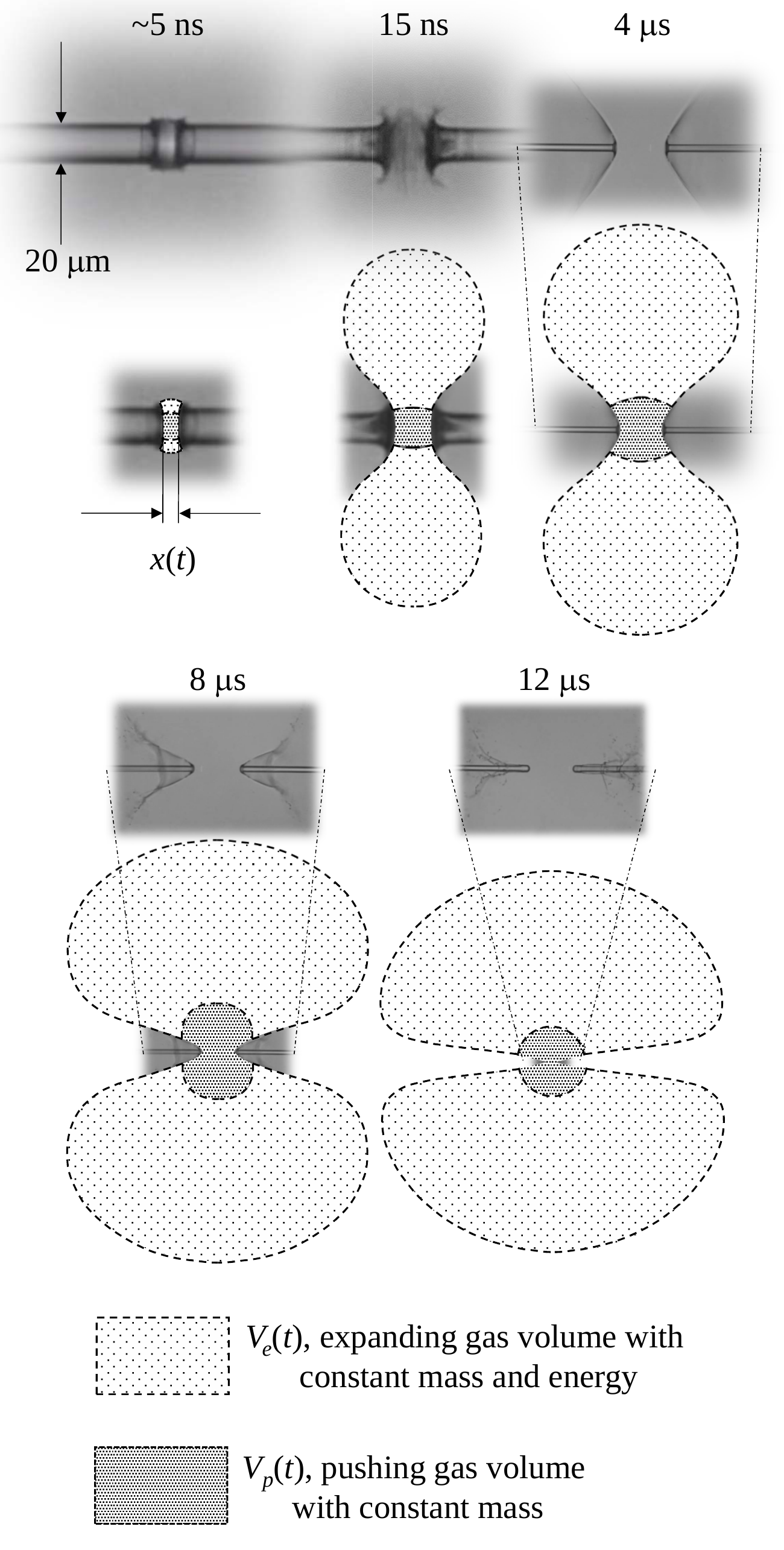}
\vspace{-3mm}
\caption{Blast induced by a strong X-ray laser pulse (photon energy 8.2 KeV, 0.75 mJ, duration 30 fs) on a liquid microjet \cite{Stan2016}. 5 ns: initial stage, where the highly compressed quasi-cylindrical pushing volume $V_p(t)$ expands against the two liquid fronts while the still nearly cylindrical expanding gas volume $V_e(t)$ does it in the radial and axial directions. Approximate illustrating shapes of $V_p$ and $V_e$ are depicted. 15 ns: $V_p$ starts expanding in the radial direction, while $V_e$ begins a doughnut-shaped mixed radial-spherical expansion. 4 $\mu$s: $V_p$ begins a mixed radial-spherical expansion. 8 $\mu$s: $V_p$ starts a nearly spherical expansion, while  $V_e$ approaches a nearly spherical expansion. Final stage ($t\gtrsim 12 \mu$s): Both $V_p$ and $V_e$ expand nearly spherically. From Stan et al. 2016 \cite{Stan2016} Suppl. Info.}
\label{fig1b}\vspace{-5mm}
\end{figure}

{\it Analysis of jet explosions.-} Figure \ref{fig1b} shows the evolution of the blast caused by a X-ray pulse (photon energy 8.2 KeV, 0.75 mJ, duration 30 fs) on a cylindrical water microjet of 20 $\mu$m discharged in vacuum. The effective beam diameter is approximately 1 $\mu$m. The energy deposited instantaneously splits the jet in two symmetrical rods whose separating fronts develop two symmetrical expanding liquid lamellas, whose energy (received from the gas) can be expressed as:
\begin{eqnarray}
\int_{t_o}^t\int_{S_p+S_i} P(\mathbf{x_p},t') \mathbf{v}_p\cdot \mathbf{n}_p\text{d} A\text{d} t= \nonumber \\
\frac{1}{2} \rho_l\frac{\pi d_j^2}{4}\int_{x_0}^x  \frac{x_t^2}{4} \text{d}x =\frac{\pi d_j^2 \rho_l}{32} \int_{t_0}^t  x_t^3\text{d} t,
\label{liquid}
\end{eqnarray}
where $x$ is the distance between the two separating liquid fronts, subscript $t$ indicates time derivative, and $\rho_l$ is the density of the liquid. $x_0$ and $t_0$ are the initial values of the gap size and time, respectively. To compute the total kinetic energy of the liquid in the two liquid lamellas in expression (\ref{liquid}), we have assumed that the liquid is radially ejected each instant at a speed $x_t/2$ due to the overpressure in $V_p$, for conservation of momentum. 

Besides, the thickness $\delta$ of the liquid layer affected by thermal conduction from the gas compared to the jet diameter $d_j$ can be estimated as $\delta/d_j\sim \left(\frac{K^2 d_j}{\rho_l c^2 E_D}\right)^{1/2}$, where $K$ and $c$ are the thermal conductivity and specific heat of the liquid. For jet sizes below 100 $\mu$m and energies $E_D$ used in SFX experiments, $\delta$  would be smaller than the molecular size, which would make the thermal energy transfer to the liquid negligible once the rapid blast takes place. This supports neglecting the internal energy gain by the liquid in (\ref{liquid}).
Thus, defining $\phi=x/l_o$, $\tau=t/t_o$ and $\Omega=V_p/(\chi V_o)$ in equation (\ref{liquid}), equation (\ref{ep1}) can be simply expressed in non-dimensional form as:
\begin{equation}
\int_{\tau_o}^\tau \phi_\tau^3 \text{d}\tau + \chi \Omega (\tau)^{1-\gamma}=\chi,
\label{ep2}
\end{equation}
with its time derivative form as:
\begin{equation}
\phi_\tau^3=\chi (\gamma-1)\Omega^{-\gamma}\Omega_\tau\Longrightarrow \phi_\tau^2=\chi (\gamma-1)\Omega^{-\gamma}\frac{\text{d} \Omega}{\text{d}\phi},
\label{ep3}
\end{equation}
where $t_o=\left(\frac{\pi \rho_l d_j^2 l_o^3}{32 E_D}\right)^{1/2}$. $l_o$ and $chi$ will be determined from dimensional arguments and maximum correlation of experimental data. On the other hand, although $\Omega(\tau)$ is an unknown variable of the problem, the mathematical structure of equation (\ref{ep3}) together with physical principles will univocally fix the asymptotic trends of $\Omega$ for both large and small times $\tau$, which is analyzed subsequently.

In SFX experiments, the energy $E_D$  is a function of the the absorption coefficient of the liquid $\alpha$, the beam profile, and the {\it volume of liquid irradiated}, assumed equal to $V_o$ (see Suppl. Material), which in turn depends on the the beam-to-jet diameter ratio $\eta=r_B/d_j$. In this type of phenomena, one may expect the existence of a minimum energy density to activate the process: the onset of electrostatic trapping of photoelectrons \cite{Hau-Riege2012,Stan2016}. Hence, we assume $E_D/V_o$ as the {\it effective} energy density driving the hydrodynamic process, expressed as:
\begin{equation}
\frac{E_D}{V_o}=\frac{\alpha (E_B-E_o)}{\pi r_B^2},
\label{effective}
\end{equation}
where $\alpha E_o/(\pi r_B^2)$ is the energy density at the onset of electrostatic trapping for a given pulse length (time) and x-ray fluence \cite{Hau-Riege2012}.

{\it Asymptotic behavior  of the pushing gas volume $\Omega$ for small and large times $\tau$.-} In the very initial stage of the blast, when the absorbed photon energy has been in part thermalized and the partial neutralization and ion excess after electron ablation has stabilized, both the expanding and pushing volumes $V_e$ and $V_p$ produce an enormous push against the two liquid fronts perpendicular to the jet axis forming the gap (figure \ref{fig1b}, time $t=$ 5 ns). The geometry of both $V_e$ and $V_p$ remains nearly cylindrical for a while, especially that of $V_p$. In these circumstances, the non-dimensional form of the pushing volume should scale as $\Omega\sim \phi$ since one should expect that its expansion would proceed predominantly in the axial direction. This occurs because the expanding volume should push against a radially expanding layer of liquid that grows at a speed $\phi_\tau$ comparable to that of the opening gap, thus preventing a radial expansion of the pushing volume $\Omega$. Therefore, assuming that $\phi\sim \tau^{\alpha_0}$, using equation (\ref{ep3}) one should have:
\begin{equation}
\tau ^{2(\alpha_0-1)} \sim \tau^{-\gamma \alpha_0}\Longrightarrow \alpha_0=\frac{2}{2+\gamma}.
\label{OT1}
\end{equation}
On the other hand, in the last stages of the blast just before the gas pressure becomes so small as to let the surface tension of the liquid produce its own retraction of the two liquid fronts (figure \ref{fig1b}, $t\gtrsim 12\,\, \mu$s), one should expect that both the expanding and pushing volumes would expand predominantly in the radial direction, which allows a self-similar solution like the one early analyzed by Wedemeyer \cite{Wedemeyer1965}. This solution yields the following radial distribution of the total energy inside the gas sphere (which includes both $V_e$ and $V_p$):
\begin{eqnarray}
\frac{P}{\gamma-1}+\rho_g \frac{v^2}{2}=P_o\left(\frac{(B t^{-1})^{3\gamma}}{\gamma-1} \left(1-\xi^2\right)^{\frac{\gamma}{\gamma-1}}+ \right. \nonumber \\ \left. \frac{\gamma}{2B^2}\left(\frac{B}{t}\right)^3\left(1-\xi^2\right)^{\frac{1}{\gamma-1}}\xi^2\right),
\end{eqnarray}
where $B=\frac{3^{1/2}(\gamma-1)}{2}$, $\xi=\frac{\rho_o B}{\gamma P_o}\frac{r}{t}$, $\rho_o$ is the initial density of the expanding gas, and $r$ the radial spherical coordinate from the center of the gap (figure \ref{fig1b}). The fundamental conclusions from this solution are:
\begin{itemize}
\item For $\xi\ll 1$ (i.e., inside $V_p$ or $\Omega$), the kinetic energy to pressure ratio becomes as small as $\xi^2$, i.e. pressure dominates over inertia in $V_p$ as anticipated.
\item According to the self-similar nature of the solution, the position of the expanding edge of $V_p$ would correspond to a constant (small) value of $\xi$ according to mass conservation for any value $\xi\leq 1$ \cite{Wedemeyer1965}, since the gas should move with a constant speed $v=a_o \xi/B$ at that edge, while $\xi=1$ is the expanding edge of $V_e$.
\end{itemize}
From those conclusions, one should expect $\Omega\sim \tau^3$ for $\tau>>1$. Again, assuming that $\phi\sim \tau^{\alpha_1}$ and using equation (\ref{ep3}) one should have:
\begin{equation}
\tau^{3(\alpha_1-1)}\sim \tau^{-3\gamma +2}\Longrightarrow \alpha_1=\frac{5}{3}-\gamma
\end{equation}
In this limit, one would have $\Omega\sim \phi^{3/(5/3-\gamma)}$. Interestingly, that solution would demand a logarithmic evolution of the gap for perfect monoatomic gases ($\gamma=5/3$). This is the only scenario for which a logarithmic evolution is contemplated, in contrast to Stan et al. \cite{Stan2016}.

In summary, one may approximately express the evolution of the gap $x(t)=l_o \phi(\tau)$ as
\begin{equation}
\phi =\phi_o \tau^{\alpha_0}\left(1+(\tau/\tau_1)^{\delta}\right)^{(\alpha_1-\alpha_0)/\delta}
\label{approxphi}
\end{equation}
where constants $\phi_o$, $\tau_1$ and $\delta$ should be obtained either from experiments or numerical simulation. Finally, $l_o$ was expected to be proportional to $d_j$ in \cite{Stan2016}. However, defining for convenience $l_o=x_0$ and expecting $V_p(t_o)=\chi V_o\sim l_o^3$, from the definition of the initial energized volume $V_o=x_0 \frac{\pi}{4} d_j^2$ one can conveniently define
\begin{equation}
 l_o = d_j \chi^{1/2},
 \label{deflo}
\end{equation}
where the earlier introduced efficiency $\chi$ should be a function of the geometry ratio $\eta=r_B/d_j$ and the ratio of initial energy density given by (\ref{effective}) to the energy density of cohesion $\rho_l H_v$ ($\simeq 2.3$ GPa for water), given by
\begin{equation}
\Pi_v=\frac{\pi r_B^2 d_j F(\eta)\rho_l H_v}{E_D}
\end{equation}
where $H_v$ is the heat of vaporization at the temperature of the liquid. Usually, one has $\Pi_v\ll 1$ and therefore a simpler functional dependency as $\chi=\chi(\eta)$ is expected.

To verify our model, we have digitized the experimental results published in \cite{Stan2016} for twelve combinations of pulse energy and jet diameters, keeping the same liquid (water) and beam focus $r_B$ constant. The summary of parameters of the experiments can be taken from the original publication \cite{Stan2016}.
\begin{figure}[htb]
\centering
\includegraphics[width=0.44\textwidth]{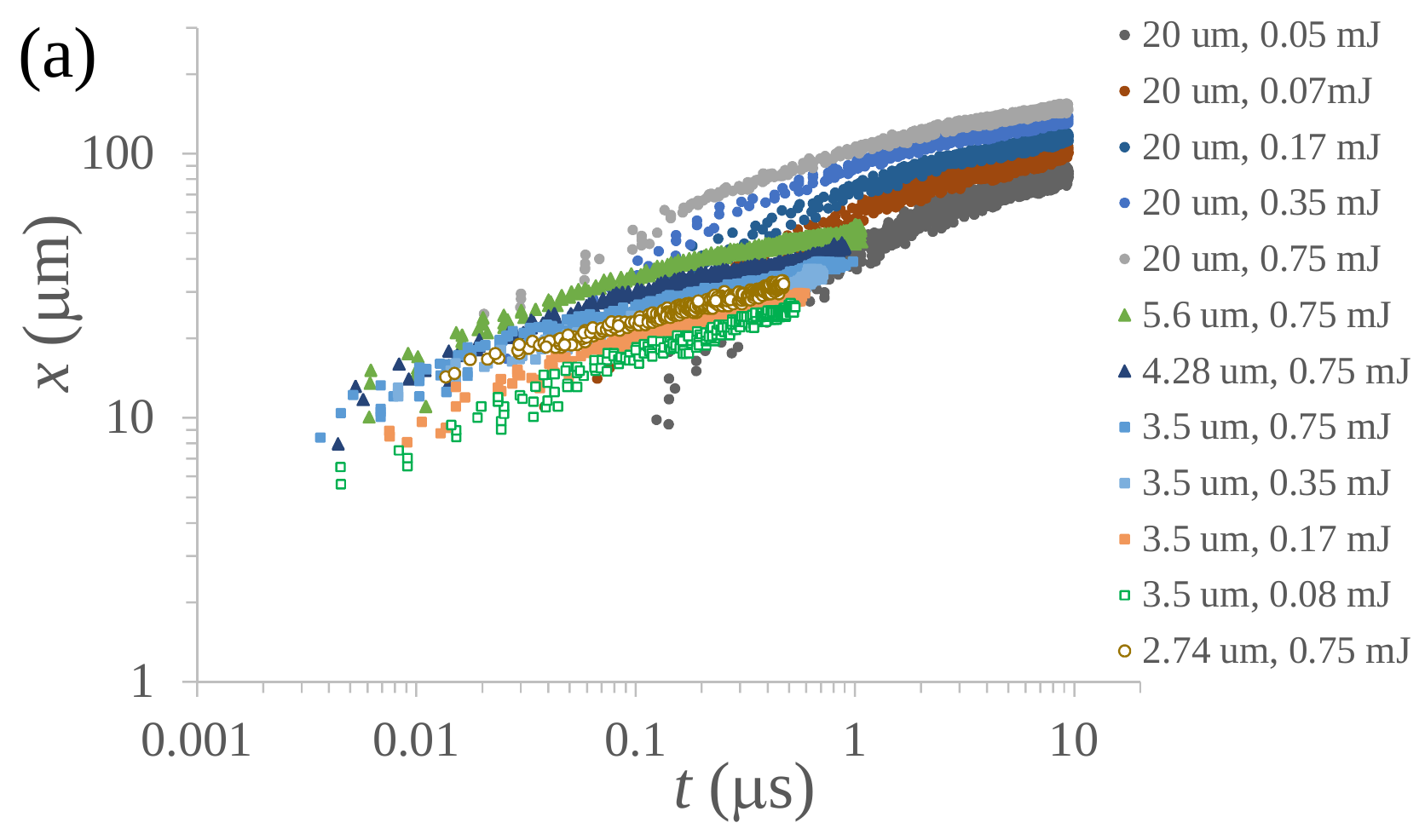}\\
\includegraphics[width=0.44\textwidth]{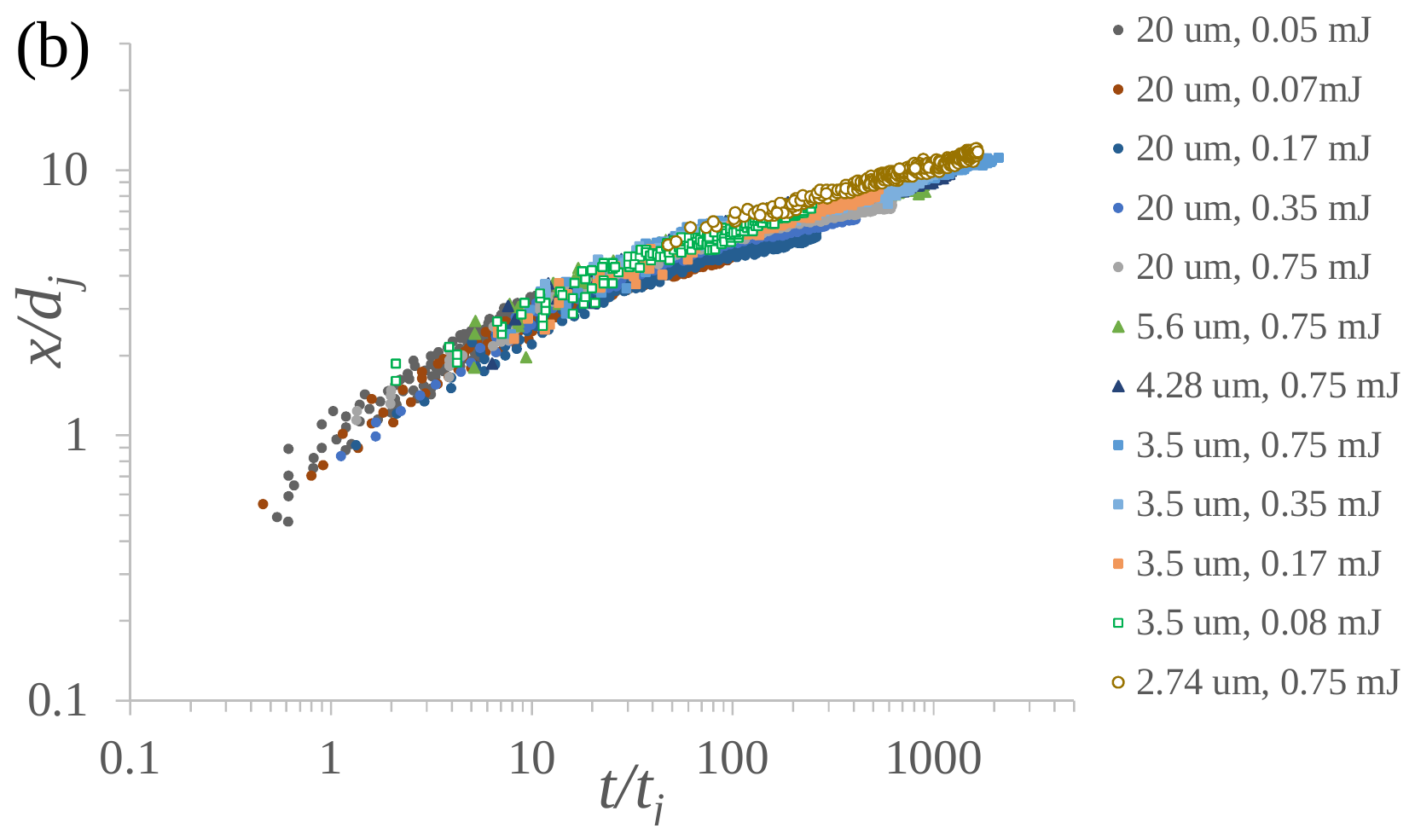}\\
\includegraphics[width=0.44\textwidth]{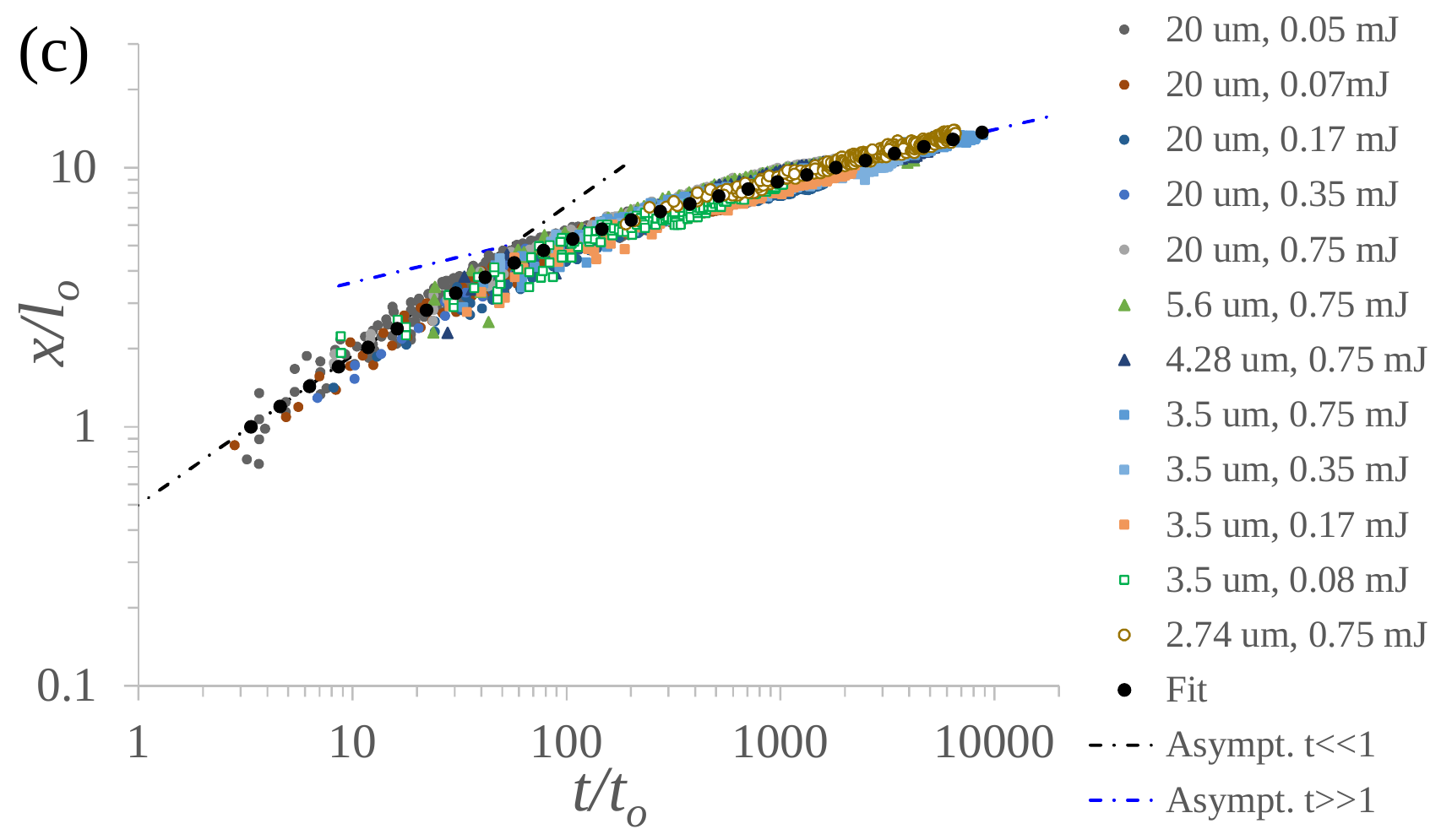}
\vspace{-3mm}
\caption{(a) Dimensional data from Stan et al. \cite{Stan2016}. (b) Sub-optimal collapse by Stan et al. \cite{Stan2016} using $t_j=\left(\rho_l d_j^5/E_D\right)^{1/2}$ and $d_j$ to make times $t$ and distances $x$ non-dimensional. (c) Optimal data collapse using $t_o=\left(\pi \rho_l d_j^2 l_o^3/(32 E_D)\right)^{1/2}$ and $l_o=d_j \eta^\beta$ to make times and distances non dimensional. The approximate analytic solution (dotted line) shows a remarkable fitting to data after optimum collapse. Dot-dashed lines: $\tau\ll 1$ (black); $\tau\gg 1$ (blue).}
\label{fig3}\vspace{-5mm}
\end{figure}
Figure \ref{fig3}(a) depicts the measurements of the gap distance $x$ as a function of time. We use the properties of water at ambient temperature ($\rho_l=1000$ kg/m$^3$, $\sigma=0.072$ N/m, $\mu_l=0.001$ Pa s$^{-1}$) for Stan's experiments. In their analysis, they readily used $d_j$ as the reference length. This election yields the collapse of data shown in figure \ref{fig3}(b) (Stan's figure S7b uses linear coordinates). However, defining $\chi=\eta^{2\beta}$ in (\ref{deflo}), one sees a significant variation of data correlation as a function of $\beta$.

We have performed a statistical correlation analysis between $\tau$ and $\phi$ data, neglecting large times for which the surface tension provokes the final retraction of the liquid fronts. We have found that both Spearman and Goodman-Kruskal correlation coefficients are maximized (see Suppl. Material) for $\beta=0.145\pm 0.002$ and $E_o=47\pm 0.5$ $\mu$J (minimum energy at the onset of electrostatic trapping). This last value would yield 60 MJ/kg for a beam radius of 0.5 $\mu$m and 31 MJ/kg for 0.7 $\mu$m, in agreement with the estimation by Stan et al. (30 MJ/kg).

Using those previous best correlation values of $\beta$ and $E_o$, in what follows we seek the best fit of the approximate function (\ref{approxphi}) to the experimental data. The minimum chi-squared logarithmic difference, shown in figure \ref{fig3}(c), is obtained for $\phi_o=0.5\pm 0.01$, $\tau_1=55\pm0.5$, and $\delta=1.5\pm 0.02$, with $\gamma=1.47\pm 0.01$ supporting the hypothesis in \cite{Stan2016} that the Gr\"uneisen coefficient for water is a nearly constant value $\Gamma=\gamma-1\simeq 0.5$ along the process. This solution is also plotted in figure \ref{fig3}(c) for reference, showing a remarkable fitting. The errors in the digitalization of data from the figures in \cite{Stan2016} are generally comparable to the errors noted in the fitting parameters. According to this solution, the maximum velocity acquired by the liquid fronts results $x_t^{(max)}/2=0.094 \eta^{\beta}\left(\frac{32 E_D}{\pi\rho_l l_o^3}\right)^{1/2}$ for $\phi=1$ and $\tau=3.35$. This yields supersonic velocities (1860 m/s) in one of the experiments ($d_j=2.74 \, \mu$m, $E_B=0.75$ mJ) only; the rest of blasts result initially (i.e. for $x_o=l_o$) subsonic for the liquid fronts. This explains why for large jets one can observe the ejection of visible strong sonic solitons (identified as shocks in \cite{Stan2016}) in the very initial stages of the explosions. Moreover, given the large numerical value of $\tau_1$ in the solution (\ref{approxphi}), the analysis of large or small values of $\tau$ should be understood as {\it compared } to $\tau_1$, obviously.

Finally, according to the definitions and from the experimental data at hand, the efficiency resulted as $\chi=(r_B/d_j)^{2\beta} = (r_B/d_j)^{0.29}$, where $r_B/d_j$ range from 0.025 to 0.2 (for $d_j$ from 20 to 2.74 $\mu$m). In other cases where $r_B>d_j$ \cite{Wiedorn2018}, further analysis would provide an extended knowledge of the efficiency $\chi$ and additional verification of physical insights here proposed. Indeed, using $r_B>d_j$ would be a way to increase the sample hit rate significantly by covering the whole jet section and allowing hits after small accidental drifts of the jet.

This work was supported by the Ministerio de Econom{\'\i}a y Competitividad (Spain), Plan Estatal 2013-2016 Retos, project DPI2016-78887-C3-1-R. Discussions with Pablo Villanueva, Claudiu Stan, Janos Hajdu, Henry Chapman, Anton Barty, and particularly with Jos\'e M. L\'opez-Herrera, Francisco Cruz-Mazo and Roberto Piriz are acknowledged. Suggestions by Pascual Riesco-Chueca are highly appreciated.



%

\end{document}


\title{Supplemental material for \\ The scaling of exploding liquid jets under intense X-ray pulses}

\author{Alfonso M. Ga{\~n\'a}n-Calvo}
\email{amgc@us.es}
\affiliation{Dept. Ingenier{\'\i}a Aerospacial y Mec{\'a}nica de Fluidos,
Universidad de Sevilla.\\
Camino de los Descubrimientos s/n 41092, Spain.}

\date{\today}

\maketitle

\appendix
\section{Control volumes involved in the formulation of blasts, and compact formulation of the conservation of energy}

Initially, a large amount of energy is injected in a partial mass $M_g$ of a deformable object with total mass $M$. $V_g(t)$ is the volume occupied by $M_g$ at each instant, and $V_D(t)$ is the volume occupied by $M$ excluding $M_g$. Hence, volumes $V_p(t)$ and $V_e(t)$ used in the formulation of the dynamics of a blast in vacuum, with partial contact with a deformable object can be formally defined by (see figure 1 in main text):
\begin{enumerate}
\item Mass conservation:
\begin{equation}
\int_{V_p+V_e}\rho_g \text{d} V=M_g,
\label{dV1}
\end{equation}
with:
\begin{equation}
\frac{\text{d}}{\text{d} t} \int_{V_p}\rho_g\text{d} V=0\quad \text{and}\quad \frac{\text{d}}{\text{d} t} \int_{V_e}\rho_g\text{d} V=0.
\label{dV2}
\end{equation}
\item Conservation of energy:
\begin{equation}
\int_{V_p+V_e+V_D}\rho\left(e + \mathbf{v}^2/2\right)\text{d} V=E_D,
\label{dV3}
\end{equation}
with
\begin{equation}
\frac{\text{d}}{\text{d} t} \int_{V_e}\rho_g\left(e + \mathbf{v}^2/2\right)\text{d} V=-\int_{S_e+S_s+S_i}P \mathbf{v}\cdot \mathbf{n}\text{d} A=0.
\label{dV4}
\end{equation}
\end{enumerate}
Here, $e$ stands for the internal energy plus the electric potential per unit mass wherever applicable, $P$ is the pressure, $\rho$ the mass density, and subindexes $l$ or $g$ stand for the deformable condensed phase or the gas phase, respectively). $\mathbf{n}$ is the outer normal at the bounding surface of $V_e$ made of the surface expanding into vacuum $S_e$, the one shared with the pushing volume $S_i$, and the one in contact with $M$, $S_s$. Equations (\ref{dV1})-(\ref{dV4}) provide a univocal definition of $V_p$ and $V_e$. Note that these definitions keep the continuity of variables and their derivatives at the surface $S_i(t)$ unaffected. In other words, these definitions do not impose any artificial boundary condition that may alter the natural evolution of the process: they just split the total gas volume for convenience into two fluid volumes that co-evolve with the same laws as the ensemble, yet having some crucial differences that allow useful simplifications of the global energy conservation equation involving the deformable object. A crucial aspect here is that the total work made by the pressure on the boundaries of $V_e$ is zero by definition, i.e.
$
-\int_{S_e+S_s+S_i}P \mathbf{v}\cdot \mathbf{n}\text{d} A=0.
$
This implies that while the pressure at the fluid interface $S_i$ is high and the product $\mathbf{v}\cdot \mathbf{n}$ is {\sl negative} at $S_i$, that product is {\sl positive} at $S_s$, such that both $-\int_{S_i}P \mathbf{v}\cdot \mathbf{n}\text{d} A$ and $-\int_{S_s}P\mathbf{v}\cdot \mathbf{n}\text{d} A$ cancel out.

Hence, one does not need to study the evolution of $V_e(t)$ and the analysis is greatly simplified. In fact, the conservation of energy for the deformable object can be written as:
\begin{equation}
\frac{\text{d}}{\text{d} t} \int_{V_D}\rho_l\left(e + \mathbf{v}^2/2\right)\text{d} V=-\int_{S_D+S_p+S_s}P \mathbf{v}\cdot \mathbf{n}_D\text{d} A
\label{eqD}
\end{equation}
where $\mathbf{n}_D$ is the outer normal on the boundaries of $V_D$. For conservation of energy, by virtue of (\ref{dV3}) and (\ref{dV4}) one has:
\begin{equation}
\frac{\text{d}}{\text{d} t} \int_{V_D}\rho_l\left(e + \mathbf{v}^2/2\right)\text{d} V+\frac{\text{d}}{\text{d} t} \int_{V_p}\rho_g\left(e + \mathbf{v}^2/2\right)\text{d} V=0
\label{eqi}
\end{equation}
Now, since $P=0$ at $S_e(t)$, and $-\int_{S_e+S_s+S_i}P \mathbf{v}\cdot \mathbf{n}\text{d} A=0$, also one has:
\begin{equation}
-\int_{S_s}P \mathbf{v}\cdot \mathbf{n}\text{d} A=\int_{S_s}P \mathbf{v}\cdot \mathbf{n}_D\text{d} A=\int_{S_i}P \mathbf{v}\cdot \mathbf{n}\text{d} A,
\label{eqP}
\end{equation}
where $\mathbf{n}$ is the outer normal on the boundaries of $V_e(t)$. Since $P=0$ at $S_D(t) $, using (\ref{eqD})-(\ref{eqP}) one can write:
\begin{equation}
-\int_{S_D+S_p+S_s}P \mathbf{v}\cdot \mathbf{n}_D\text{d} A=\int_{S_p+S_i}P \mathbf{v}\cdot \mathbf{n}\text{d} A= -\frac{\text{d}}{\text{d} t} \int_{V_p}\rho_g\left(e + \mathbf{v}^2/2\right)\text{d} V,
\end{equation}
which integrated from $t=t_o$ to an arbitrary time $t$ gives the energy balance for the pushing volume $V_p(t)$:
\begin{equation}
\left. \int_{V_p}\rho_g\left(e + \mathbf{v}^2/2\right)\text{d} V\right|_{t_o}^t +\int_{t_o}^t \int_{S_p+S_i}P \mathbf{v}\cdot \mathbf{n}\text{d} A\,\text{d} t' =0.
\label{pushing}
\end{equation}
Therefore, by virtue of (\ref{eqi}), the energy received by the liquid can be expressed as:
\begin{equation}
\left. \int_{V_D}\rho_l\left(e + \mathbf{v}^2/2\right)\text{d} V\right|_{t_o}^t= \int_{t_o}^t \int_{S_p+S_i}P \mathbf{v}\cdot \mathbf{n}\text{d} A\,\text{d} t' .
\end{equation}
Finally, it can be shown (se main text) that the kinetic energy of the gas in $V_p$ should be negligible compared to the internal energy, and therefore one can write:
\begin{equation}
\left. \int_{V_p}\rho_g\left(e + \mathbf{v}^2/2\right)\text{d} V\right|_{t_o}^t=\frac{P_oV_p(t)^{1-\gamma}}{(\gamma -1)V_o^\gamma}-\chi E_D,
\end{equation}
or, equivalently, from (\ref{pushing}):
\begin{equation}
\frac{P_oV_p(t)^{1-\gamma}}{(\gamma -1)V_o^\gamma}+\int_{t_o}^t \int_{S_p+S_i}P \mathbf{v}\cdot \mathbf{n}\text{d} A\,\text{d} t' =\chi E_D
\end{equation}
which is the compact formulation of the conservation of energy used in the main text, equation (2). Here, $\chi$ is the efficiency of the blast defined as the initial energy of $V_p(t_o)$ over the energy $E_D$ of the entire initial volume $V_o=V_p(t_o)+V_e(t_o)$. Given that (i) both volumes $V_p(t_o)$ and $V_e(t_o)$ share the same initial conditions of pressure, energy and (zero) velocity at $t=t_o$, and that (ii) pressures can be considered nearly homogeneous at $V_p(t)$, then one also has:
\begin{equation}
\chi=V_p(t_o)/V_o
\end{equation}

\section{Irradiated volume and energy deposited by the X-ray pulse}

The beam profile, or the distribution of the beam energy intensity $I(r)$ is a feature of each device. Without restriction on the generality of the analysis procedure, we could make the highly simplified assumption that the distribution is Gaussian and axisymmetric, as
 \begin{equation}
 I(r)=E_B\text{e}^{\frac{-r^2}{2 r_B^2}}/(2\pi r_B^2),
 \end{equation}
 where $E_B$ is the pulse beam energy, and $r_B$ is the beam radius inside which the energy is $(1-\text{e}^{-1/2})E_B$. Hence, the energy deposited by the beam in the jet can be written as:
\begin{equation}
\bar{E}_D=\alpha \!\! \int_0^{\pi/2} \!\! \int_0^{\frac{d_j}{2\sin \theta}} \!\!  I(r)\!\! \left(\left(d_j/2\right)^2\!\! -\! \left(r \sin\theta\right)^2 \right)^{1/2}\!\! 8 r \text{d}r\text{d}\theta,
\label{ED1}
\end{equation}
where $\alpha=\eta_l \rho_a$ is the absorption coefficient of the liquid, and $\eta_a$ is the photon attenuation coefficient \cite{Lide2004}, a function of the photon energy of the beam for a given liquid. For water and a photon energy of 8.2 KeV, one has $\alpha=1054$ m$^{-1}$. In non-dimensional form, equation (\ref{ED1}) reads:
\begin{eqnarray}
\frac{\bar{E}_D}{\alpha d_j E_B}=\int_0^{\pi/2} \!\! \int_0^{\frac{1}{2\sin \theta}} \frac{4 x \text{e}^{\frac{-x^2}{2 \eta^2}}}{\pi \eta^2} \left(\frac{1}{4}-\left(x \sin\theta\right)^2 \right)^{1/2}\text{d}x\text{d}\theta \nonumber \\
\equiv F(\eta),\quad\quad
\end{eqnarray}
where $x=r/d_j$, and $\eta=r_B/d_j$ is the beam-to-jet diameter ratio, a fundamental parameter of the jet-beam interaction. The function $F(\eta)$ can be accurately approximated by:
\begin{equation}
F_a(\eta)=\left(1+\left(\eta/\eta_o\right)^{\delta_o}\right)^{-1/\delta_o}
\end{equation}
where a minimization of the Chi-squared difference taking $N$ points logarithmically distributed as $\eta_i=N^{(j/N)}$ (i.e., minimizing $\sum_j^N \left(\log\left(\frac{F_a(\eta_i)}{F(\eta_i)}\right)\right)^2$) yields $\eta_o=0.31203$ and $\delta_o=2.5371$ with 20 points and an average error below 5.4$\times 10^{-5}$. In summary, the fraction of the pulse energy deposited can be accurately expressed as:
\begin{equation}
\bar{E}_D=E_B \alpha d_j F_a(\eta)
\label{EEE}
\end{equation}
We can define an equivalent irradiated volume $V_o$ as that where the energy deposited is $\bar{E}_D$ and the energy density is equivalent to that of a cylinder with section $\pi r_B^2$ and length $1/\alpha$ where the whole pulse energy $E_B$ would have been absorbed. Hence, according to that definition, one has:
\begin{equation}
V_o=\frac{\bar{E}_D}{\alpha E_B/(\pi r_B^2)}
\end{equation}
or, using (\ref{EEE}),
\begin{equation}
V_o=\pi r_B^2 d_j F_a(\eta)
\label{EEQ}
\end{equation}
Assuming the existence of a minimum energy density $\alpha E_o/(\pi r_B^2)$ to activate the process, by virtue of (\ref{EEQ}) the equivalent irradiated volume $V_o$ is also that where the {\it effective} energy deposited is $E_D$ and the energy density is the same as that of a cylinder with section $\pi r_B^2$ and length $1/\alpha$ where an {\it effective} pulse energy $E_B-E_o$ is injected, i.e.
\begin{equation}
V_o=\frac{E_D}{\alpha (E_B-E_o)/(\pi r_B^2)}=\frac{\bar{E}_D}{\alpha E_B/(\pi r_B^2)}
\end{equation}

\section{Optimal correlation analysis of data}

\begin{figure}[htb]
\centering
\includegraphics[width=0.95\textwidth]{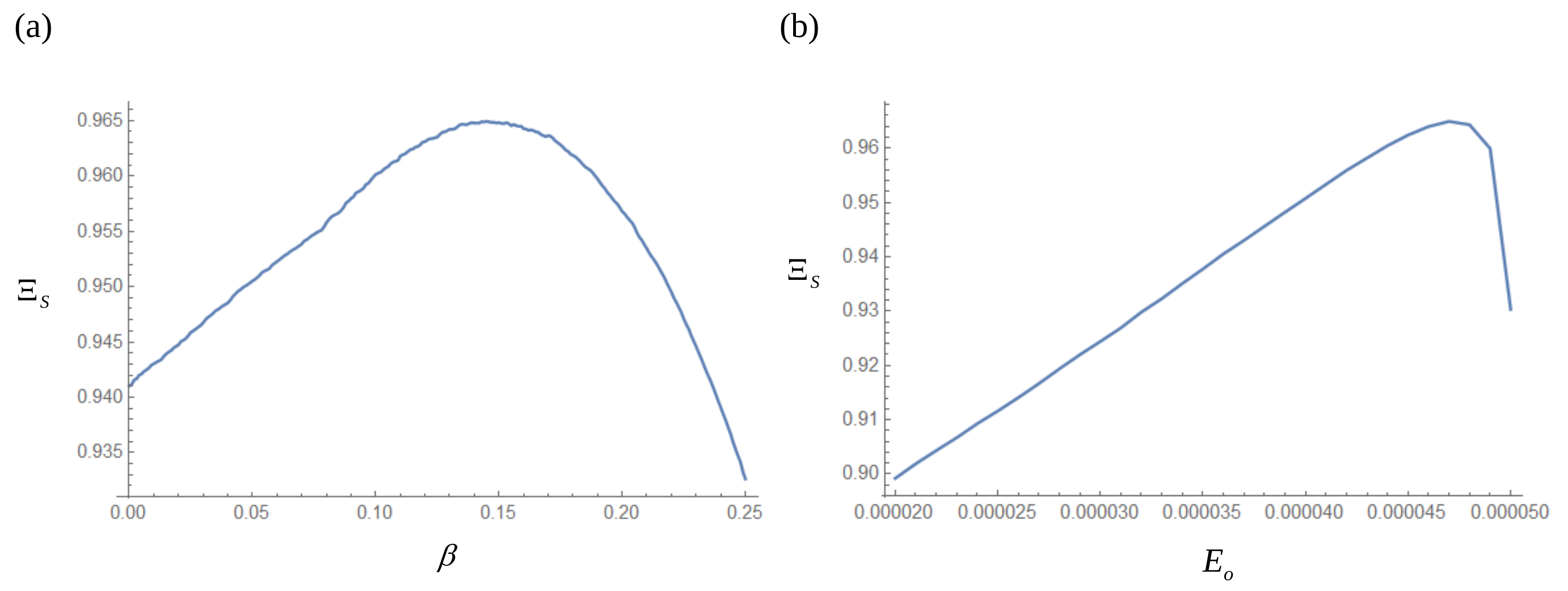}\\
\vspace{-3mm}
\caption{Spearman correlation coefficient $\Xi_S$ as a function of the two variables $\beta$ and $E_o$. (a) Section of the surface for $E_o=4.7\times 10^{-5}$ J. (maximum at $\beta=0.145$); (b) Section of the surface for $\beta=0.145$ (maximum at $E_o=4.7\times 10^{-5}$ J).}
\label{fig1SM}
\end{figure}

\begin{figure}[htb]
\centering
\includegraphics[width=0.95\textwidth]{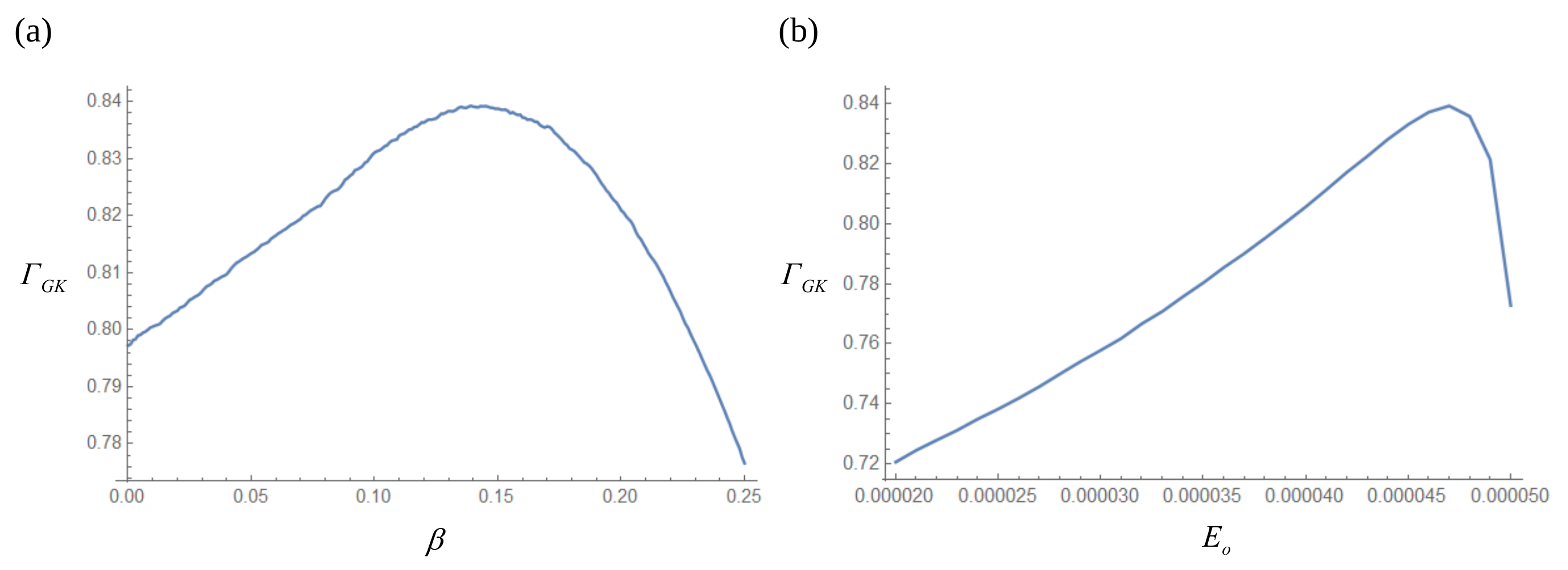}\\
\vspace{-3mm}
\caption{Goodman-Kruskal $\Gamma_{GK}$ as a function of the two variables $\beta$ and $E_o$. (a) Section of the surface for $E_o=4.7\times 10^{-5}$ J. (maximum at $\beta=0.145$); (b) Section of the surface for $\beta=0.145$ (maximum at $E_o=4.7\times 10^{-5}$ J).}
\label{fig2SM}
\end{figure}

We have performed a double parameter optimization of both Spearman $\Xi_S$ \cite{Spearman1904} and Goodman-Kruskal $\Gamma_{GK}$ \cite{Goodman1954,Goodman1959,Goodman1963,Goodman1972} correlation coefficients for the non dimensional data $\tau$ and $\phi$ using the statistical package of Mathematica$^\circledR$. Figures \ref{fig1SM} and \ref{fig2SM} show the sections of the surfaces  $\Xi_S=\Xi_S(\beta,E_o)$ and $\Gamma_{GK}=\Gamma_{GK}(\beta,E_o)$, respectively, for constant values of either $E_o$ or $\beta$ where maxima are found, to show how the functions behave around these maxima. Observe the evident existence of a single conspicuous maximum for both  $\Xi_S=\Xi_S(\beta,E_o)$ and $\Gamma_{GK}=\Gamma_{GK}(\beta,E_o)$.

%